\documentstyle[sprocl]{article}

\def\be{\begin{equation}}
\def\ee{\end{equation}}
\def\bea{\begin{eqnarray}}
\def\eea{\end{eqnarray}}

\begin{document}


\title{Black Hole Accretion in Transient X-Ray Binaries}

\author{Kristen Menou \footnote{Chandra Fellow}}

\address{Princeton University, Department of
Astrophysical Sciences, Princeton, NJ 08544, USA. ~E-mail:
kristen@astro.princeton.edu}


\maketitle\abstracts{Recent work on the modes of accretion onto black
holes (BHs) in Soft X-Ray Transients (SXTs) is reviewed, with an
emphasis on uncertainties affecting models of accretion during
quiescence (inner hot flow, outer thin disk). Various interpretations
of the quiescent X-ray luminosity difference between systems
containing neutron stars (NSs) and systems containing BH candidates
are also summarized. A new scenario, which does not require BH
candidates to possess an event horizon, is presented here. This
scenario may be ruled out in the future, from detailed X-ray
spectroscopic diagnostics or from the absence of type I X-ray bursts
in systems containing BH candidates.  }

\section{Introduction}

One of the main interests for studying accretion in close binary star
systems is that we know better in these systems than anywhere else the
general conditions under which accretion proceeds (e.g.  binary size
and geometry, donor and accretor masses, etc...). Many such close
binary systems are transient: they semi-regularly experience large
amplitude outbursts followed by long periods of quiescence. Depending
on the nature of the compact object, these systems are called Dwarf
Novae (DN; for a white dwarf accretor; see Warner 1995 for a review)
or Soft X-ray Transients (SXTs; for a neutron star or black hole
accretor; see Lewin, van Paradijs \& van den Heuvel 1995 for a
review).

Although this contribution is focused on the nature of accretion in BH
SXTs (mostly during quiescence), reference will also be made to NS
SXTs and DN. The similarities between these three classes of systems
are strong, so that one can hope to learn about the compact object
from their subtle differences. In \S2, I review the motivation for a
two-component accretion flow structure (inner hot flow, outer thin
disk) in quiescent BH SXTs. In \S3, I discuss various uncertainties
affecting models of these two components of the accretion flow. In
\S4, I review the observational status and theoretical interpretations
of the quiescent X-ray luminosity difference between systems
containing BHs and those containing NSs. In \S5, I present an
alternative interpretation of this luminosity difference that does not
require BH candidates to possess an event horizon.

\section{Accretion Flow Structure}

The gas accreted onto compact objects in close binary systems is
transferred via Roche-lobe overflow by a main-sequence or sub-giant
companion. The large specific angular momentum of the gas naturally
leads to the conclusion that a thin accretion disk forms in the system
(the gas being allowed to radiatively lose energy much more
efficiently than momentum when cooling is efficient; see, e.g., Frank,
King \& Raine 1992 for a review). The presence of thin accretion disks
during outburst in transient close binaries has been observationally
confirmed by detailed eclipse maps in DN (Horne 1993) and X-ray
spectroscopic diagnostics in SXTs (Tanaka \& Shibazaki 1996). The
situation in quiescence appears more complicated, however.

According to the Disk Instability Model (DIM), the outbursts of
transient close binaries are triggered by a thermal-viscous
instability occurring when hydrogen becomes partially ionized
somewhere in the disk (Meyer \& Meyer-Hofmeister 1981; Cannizzo 1993;
Lasota 2001a). Lasota (1996) pointed out that if fully-extended,
quiescent disks in SXTs were accurately described by the DIM,
accretion onto the compact object should proceed at a rate $\ll
10^{10}$~g~s$^{-1}$, i.e. many orders of magnitude smaller than
inferred from X-ray observations. In addition, at such a low rate, a
thin accretion disk is not expected to emit a substantial amount of
X-rays (Narayan, McClintock \& Yi 1996). This has led to the general
belief that there cannot be a disk extending all the way down to the
compact object in quiescent SXTs, as would be expected in the
DIM. Wheeler (1996) also pointed out that the optical light from
quiescent BH SXTs (excluding the contribution from the companion star;
see Narayan et al. 1996) corresponds to black body temperatures in
excess of $10^4$~K, which is too hot for the neutral disk expected in
the DIM. Nonetheless, the presence of a disk in quiescence appears
required to explain the double-peaked emission lines observed
(e.g. Orosz et al. 1994) and the outbursts experienced by SXTs.

An attractive solution to this problem is to postulate that the thin
disk is present only in the outer regions of the accretion flow, while
it is replaced by a hot, X-ray emitting flow in the vicinity of the
BH. This possibility was first proposed by Narayan et al.  (1996) and
further developed by Esin, McClintock \& Narayan 1997 (see also
Narayan 1996; Lasota, Narayan \& Yi 1996; Narayan, Barret \&
McClintock 1997; Hameury et al. 1997). These specific models, in which
the inner hot flow is assumed to be an advection-dominated accretion
flow (ADAF), have been rather successful at explaining the spectral
properties of BH SXTs in various spectral states (Esin et al. 1998).

Strong observational support for the presence of a hot flow in the
inner regions of BH SXTs during low-luminosity spectral states exists
for the unusual BH SXT labeled XTE~J1118+480 (see McClintock et
al. 2001a for a mass function determination). This system is located
high above the Galactic plane, so that for the first time EUV and soft
X-ray spectral measurements could be obtained for a BH SXT (given the
small extinction to the source; Garcia et al. 2000). These
spectroscopic data were obtained as part of an extensive
multiwavelength campaign during the low/hard-state outburst that led
to the discovery of this system (Hynes et al. 2000; McClintock et
al. 2001b and references therein). The low-state spectral energy
distribution of XTE~J1118+480 strongly suggests the presence of a
thermal component peaking at $\sim 50$~eV, in addition to the hard
power law seen at higher energies (McClintock et al. 2001b). As
explicitly shown by Esin et al. (2001), these data are best
interpreted as indicating that the thin disk in this system
(responsible for the thermal emission) is truncated at several tens of
Schwarzschild radii from the BH, where it is replaced by a hot
accretion flow (responsible for the power law emission observed). It
is difficult to interpret the thermal emission observed as power-law
emission from a corona that has been reprocessed by a fully extended,
underlying disk (because the thermal component is apparently more
energetic than the power-law component; Esin et al. 2001).

\section{Uncertainties}

Based on the successful spectral modeling of Esin et al. (1998) and
the evidence in XTE~J1118+480 (as well as other arguments discussed in
\S2), there are good reasons to believe that the inner regions of thin
disks in quiescent BH SXTs are replaced by hot accretion flows. Many
important components of this scenario are poorly constrained,
however. A major uncertainty comes from our lack of understanding of
the nature of the transition between the two components of the
accretion flow. Although several mechanisms for the ``evaporation'' of
the disk into a hot flow have been explored in the literature (Meyer
\& Meyer-Hofmeister 1994; Honma 1996; de Kool \& Wickramasinghe 1999;
Rozanska \& Czerny 2000; Manmoto \& Kato 2000; Spruit \& Deufel 2001;
Ball, Narayan \& Quataert 2001), none of these theories provide
reliable predictions that can be put to the observational test. In
what follows, the discussion will be focused on additional
uncertainties affecting models of the inner hot flow and the outer
thin disk separately.

\subsection{Inner Hot Flow}

The work on hot accretion flows in recent years has focused on
self-similar, analytical solutions to the problem of accretion onto a
BH (but see also the work by Ichimaru 1977; Rees et al. 1982;
Abramowicz et al. 1995).

Narayan \& Yi (1994; 1995a,b) derived a self-similar,
advection-dominated accretion flow (ADAF) solution in which nearly all
the gravitational potential energy released during accretion is
advected by the flow rather than efficiently radiated locally by the
gas (as is the case for thin disk accretion). This advection property
requires preferential viscous heating of the ions over the electrons
and a two-temperature structure. The latter is allowed if energy
transfer between the ion and electron populations is rather
inefficient, as is the case if only Coulomb collisions operate. A
detailed analysis of the energetics of turbulence and the various
dissipation mechanisms involved in this type of flow suggests that
preferential ion heating can be achieved only for significantly
sub-equipartition magnetic fields (Gruzinov 1998; Quataert 1998;
Quataert \& Gruzinov 1999). As initially pointed out by Narayan \& Yi,
two complications with the ADAF solution are that the flow has a
positive Bernoulli energy constant (indicating that the gas may be
subject to hydrodynamical outflows because it is technically unbound)
and is unstable to convection in the radial direction (while the
effects of convection on the ADAF structure are neglected).

Blandford \& Begelman (1999) explored further the possibility that the
hot flow is subject to outflows. They constructed a family of
advection-dominated inflow-outflow solutions (ADIOS) in which mass,
energy and angular momentum are gradually lost to an outflow as part
of the gas is being accreted. They suggested that solutions with
negative Bernoulli energy constant, which can be obtained under strong
outflow assumptions, are more viable than the initial ADAF solution
(see Abramowicz, Lasota \& Igumenshchev 2000 for a different view).

Alternatively, Narayan, Igumenshchev \& Abramowicz (2000) and Quataert
\& Gruzinov (2000) explored the role of convection in determining the
structure and properties of the hot flow.  These authors proposed a
convection-dominated accretion flow (CDAF) solution in which radial
convection is postulated to dominate over viscous processes to
determine the sign of angular momentum transport in the flow. The
properties of CDAFs are quite different from those of ADAFs, in
particular the presence of an additional convection-driven, outward
energy flux that may be at the origin of an outflow launched from the
outer regions of the hot flow. An interesting feature of the CDAF
solution is that it does not require preferential ion heating like the
ADAF solution does (Quataert \& Narayan 1999; Ball et al. 2001).

While the temperature structure of ADAFs, ADIOS and CDAFs is close to
virial in all three cases, their density structure vary widely: $\rho
\propto R^{-3/2}$ in an ADAF, $\rho \propto R^{-1/2}$ in a CDAF and a
$\rho$ profile that depends on the wind parameters in an ADIOS. Given
an accretion geometry and an accretion rate at which the truncated
disk feeds the hot flow in a quiescent BH SXT, the X-ray emission
properties will therefore be largely different depending on the exact
nature of the hot flow. The ADAF solution may not be favored because
of the neglect of radial convection, the apparent susceptibility to
hydrodynamical outflows and the possible difficulty to preferentially
heat ions. It is unclear, however, whether the hot flow in quiescent
BH SXTs should have an ADIOS-like structure, a CDAF-like structure or
none of the above.

Given this uncertain situation, the interest for numerical
investigations of the structure of radiatively inefficient hot flows
has grown considerably during the last few years. A large number of 2D
and 3D hydrodynamical simulations (Igumenshchev, Chen \& Abramowicz
1996; Igumenshchev \& Abramowicz 1999; 2000; Stone; Pringle \&
Begelman 1999; Igumenshchev, Abramowicz \& Narayan 2000) have
indicated that hot flows with a small value of the viscosity parameter
$\alpha$ ($<0.1$ or so) are subject to large-scale convective motions
(in better agreement with CDAF theory), while hot flows with a large
$\alpha$ value ($>0.1$ or so) are subject to strong outflows (in
better agreement with ADIOS theory). An obvious shortcoming of all
these studies is that an ad-hoc Shakura-Sunyaev-type viscosity has to
be included in the Navier-Stokes equations solved to mimic the
``viscous'' interactions that are thought to be in fact of magnetic
origin in this hot plasma (Balbus \& Hawley 1991; 1998)

Results of global 3D magneto-hydrodynamical (MHD) numerical
simulations of hot flows have appeared only very recently because of
the large numerical power required. Hawley, Balbus \& Stone (2001)
present such a calculation, with strong outflow properties. In
addition, these authors argue that treating convection in the hot flow
separately from angular momentum transport, as is done in CDAF
analytical theories, is incorrect. On the other hand, Igumenshchev \&
Narayan (2001) present a 3D MHD numerical simulation of hot, spherical
accretion (gas with zero net angular momentum) in which convective
motions play an important role. The apparent discrepancy between these
results may be caused by differences in the physical problems
addressed or perhaps different numerical implementations
(e.g. allowing or not for explicit magnetic energy dissipation in the
equations solved), but it is clear at this point that no consensus has
emerged on the structure and properties to expect for hot
flows. Additional numerical simulations will likely resolve this issue
in the future.

\subsection{Outer Thin Disk}

The structure and properties of outer thin disks in quiescent BH SXTs
are also subject to considerable uncertainties, mostly because of our
ignorance of the nature of viscosity in this case. These uncertainties
are large enough that we are currently unable to predict the rate at
which the disk feeds the inner hot flow in these systems (given a mass
transfer rate). Although the DIM makes definite predictions for the
accretion rate as a function of radius to expect in a quiescent disk
($\dot M \propto R^{2.5}$ or so), the assumption that the bulk of the disk
accretes mass with a uniform efficiency of angular momentum transport
(expressed as a constant value $\alpha$ parameter in the DIM) may not
be valid.

While the transport in ionized disks during outburst is probably due
to MHD turbulence resulting from the non-linear development of the
Magneto-Rotational Instability (MRI; Balbus \& Hawley 1991; 1998),
Gammie \& Menou (1998) pointed out that quiescent disks are so neutral
that resistive diffusion is important and MHD turbulence may not be
sustained during this phase. Menou (2000) reaffirmed by using the MHD
numerical simulations of resistive disks of Fleming, Stone \& Hawley
(2000) in more detailed calculations.

Recently, however, Wardle (1999) and Balbus \& Terquem (2001) pointed
out that the role of Hall terms had generally been ignored in studies
of weakly-ionized disks.  Although Hall terms are strongest in
low-density environments (hence the discussion for T-Tauri disks),
they are not negligibly small in the quiescent disks of close binary
systems. Using the scaling derived by Balbus \& Terquem (2001; their
Eq.~25) and assuming a reasonable sub-equipartition magnetic field,
one finds that the resistive diffusion term dominates over Hall
effects for typical densities $\sim 10^{-6}$~g~cm$^{-3}$ found in
quiescent disks. Nonetheless, Hall terms constitute an additional
non-ideal MHD effect that will have to be included in future
discussions of the level of MHD turbulence to expect in quiescent
disks in transient close binaries.\footnote{Balbus \& Terquem note how
difficult it is to analytically predict the influence of Hall terms
because of their potentially stabilizing {\it and} destabilizing
effects.}

Menou (2000) argued that, in the absence of transport by MHD
turbulence during quiescence, another ``viscosity'' mechanism could
then drive accretion (perhaps spiral density waves induced in the disk
via tidal interaction with the companion star; Spruit 1987).  Even in
the absence of an alternative viscosity mechanism operating during
quiescence, it is still possible for accretion to proceed via
MHD-turbulent, X-ray ionized layers at the disk surface. This
layered-accretion scenario, first considered by Gammie (1996) for
T-Tauri disks (ionized by cosmic rays), has been discussed in detail
by Menou (2001) for quiescent disks in DN. The feasibility of layered
accretion has been numerically demonstrated by Fleming \& Stone
(2002).

It is unclear, however, how relevant the layered accretion model is
for quiescent disks in SXTs, because the origin and geometry of the
ionizing X-rays are not well understood in these systems (for
instance, the low-level of X-ray emission in quiescent BH SXTs may not
allow the surface layers to be sufficiently ionized for MHD turbulence
to operate). Nevertheless, the possibility that accretion proceeds via
surface layers in quiescent DN rather than in the bulk of the disk as
assumed in the DIM shows that our ignorance of the nature of viscosity
in quiescent disks results in large uncertainties on the structure and
properties of these disks. In particular, layered accretion predicts
$\dot M \propto R$ in quiescent disks, as compared to $\dot M \propto
R^{2.5}$ or so for the DIM (Menou 2001). Therefore, the rate at which
the quiescent disk feeds the hot flow in BH SXTs is currently
unpredictable. This situation is unlikely to change until we better
understand the mechanism responsible for accretion in quiescent disks.

\section{Black Hole -- Neutron Star Luminosity Difference}

One of the most interesting properties of BH SXTs is their X-ray
faintness during quiescence, as compared to NS SXTs. The existence of
this difference was first pointed out by Narayan, Garcia \& McClintock
(1997) and Garcia et al. (1997), when they compared systems
experiencing type I X-ray bursts (hence containing NSs) to those with
mass functions in excess of $2-3 M_\odot$ (presumably containing
BHs). Recent deep {\it Chandra} observations allowed the detection of
several faint quiescent BH SXTs, thus confirming with better
statistics that quiescent BH systems are about 2 orders of magnitude
fainter than their NS equivalents (see Garcia et al. 2001 for the
latest data). Since the discovery of this luminosity difference,
various interpretations of the observational data have been proposed
and debated, as described below.

\subsection{Accretion Scenarios}

Narayan et al. (1997) interpreted the data on quiescent SXTs as
evidence for event horizons in systems containing BH
candidates. Indeed, assuming that accretion proceeds via an ADAF in
the inner regions of quiescent BH SXTs (\S2), a large fraction of the
gravitational potential energy released during the accretion process
may be stored as thermal energy in the flow and lost through the BH
event horizon (thus leading to a very small radiative efficiency). On
the contrary, independent of the structure of the accretion flow, one
would expect a large radiative efficiency ($\sim 10-20 \%$) in
quiescent NS SXTs, from the necessity to radiate all the energy
released by accretion at the stellar surface. Assuming that accretion
proceeds at roughly the same rate in the two classes of systems, one
therefore expects quiescent BH SXTs to be less luminous than their NS
equivalents.

Menou et al. (1999) attempted to quantify the luminosity difference
expected in this accretion scenario with more detailed models. They
pointed out the importance of comparing systems with similar orbital
periods, to guarantee comparable mass transfer rates and (presumably)
mass accretion rates (allowing a meaningful test). They also showed
that quiescent NS SXTs are much less luminous than naively expected,
in that only a small fraction ($\sim 10^{-3}$) of the mass transferred
must actually reach the NS surface. This could be achieved via the
action of an efficient ``propeller effect'' in these systems (see also
Asai et al. 1998). Menou \& McClintock (2001) attempted to further
test the ADAF+propeller scenario for quiescent NS SXTs, by using
multiwavelength spectral data available for the specific system Cen
X-4. This was not very successful in that the data do not show any
signature of the presence of an ADAF in this system. Chandler \&
Rutledge (2000) further questioned the validity of the propeller
picture by showing the near absence of X-ray pulsations when the NS
SXT Aql X-1 fades into quiescence, while it is supposed to enter the
propeller regime at that point.

Recently, Loeb, Narayan \& Raymond (2001) and Abramowicz \&
Igumenshchev (2001) suggested that the observed X-ray luminosity
difference of a factor $\sim 100$ would be better understood if one
accounts for the modified hot flow structure expected from strong
convection (CDAF theory). It is unclear how reliable these
interpretations are, however, because of various oversimplifications
(e.g. neglect of the stellar magnetic field, unproven relevance of
CDAF theory for the case of accretion onto a compact object with a
hard surface; see also discussion in Lasota 2001b). The current
situation with accretion scenarios for quiescent BH and NS SXTs
largely reflects the numerous uncertainties associated with hot flow
models, as discussed in \S3.

\subsection{Stellar Scenarios}

Bildsten \& Rutledge (2000) proposed that the quiescent X-ray emission
of BH SXTs is dominated by coronal emission from the rapidly-rotating
companion star in these systems. This possibility has been challenged
by Lasota (2000) and cannot explain the excessive X-ray emission in at
least two know quiescent BH SXTs (GRO~J0422+32 and V404~Cyg; see
Bildsten \& Rutledge 2000; Narayan, Garcia \& McClintock 2001). X-ray
spectral diagnostics also challenge this scenario in several
additional systems (Kong et al. 2001).

Brown, Bildsten \& Rutledge (1998) pointed out that the (soft) X-ray
emission of quiescent NS SXTs could be powered by NS thermal cooling,
following heating by deep crustal compression during outburst. The
X-ray luminosities of $\sim 10^{32}-10^{33}$~erg~s$^{-1}$ obtained by
Brown et al. (1998), in agreement with the observed values, were later
confirmed in more detailed calculations by Colpi et al. (2001; see
also Ushomirsky \& Rutledge 2001). The unusually low quiescent X-ray
luminosity of the ms X-ray pulsar SAX~J1808-36 (Wijnands et al. 2001a)
was expected in this scenario given the outburst properties of this
system (Brown et al. 1998).  The NS cooling interpretation is further
supported by detailed NS hydrogen atmosphere fits to the X-ray data
which require an emission region of size comparable to the entire NS
surface, as expected (Rutledge et al. 1999; 2000; 2001a,b). These
atmosphere models are successful in fitting the soft X-ray spectra of
several sources recently discovered in globular clusters as well
(Rutledge et al. 2001c; Grindlay et al. 2001). The crustal heating
scenario may also be able to explain the case of the NS in the SXT
KS~1731-260, despite the unusual outburst and cooling properties of
this system (Wijnands et al. 2001b; Rutledge et al. 2001d).

Despite successes, the crustal heating scenario is not without
difficulties. The power law component seen in the X-ray spectra of
several quiescent NS SXTs (Asai et al. 1996; Campana et al. 1998,
Rutledge et al. 2001a,b) is not explained by this scenario and
therefore requires a different origin. It is puzzling that in both Aql
X-1 and Cen X-4, the thermal and power law components have comparable
luminosities, while they should be unrelated according to the crustal
heating scenario. This coincidence may be more easily explained in an
accretion scenario (which is not incompatible with
hydrogen-atmosphere-type emission).

Variability properties of quiescent NS SXTs may also challenge the
crustal heating model. Rutledge et al. (2001a) attributed the
long-term variability of Cen X-4 during quiescence to a variation in
the power-law component only. Campana et al. (1997) reported
significant variability of the quiescent X-ray emission of Cen X-4
(over a period of several days), but it is unclear whether this
variability can be simply attributed to the contribution of the power
law component in the ROSAT-HRI soft X-ray band. Ushomirsky \& Rutledge
(2001) suggest that transient accretion events could lead to specific
variability patterns of the NS thermal emission in the crustal heating
scenario. Nonetheless, Rutledge et al. (2001b) acknowledge that the
long-term variability of the thermal component that they observed in
Aql X-1 during quiescence is not easily explained by the crustal
heating scenario.

\subsection{Other Scenarios}

Other interpretations of the X-ray emission from quiescent SXTs have
been proposed, including the radio-pulsar interpretation of Campana \&
Stella (2000) for NS SXTs. These alternatives are critically reviewed
by Narayan et al. (2001).

\section{Evidence for Event Horizons?}

Whether the quiescent X-ray emission of SXTs is powered by accretion
or NS cooling, the scenarios discussed in the previous section
attribute, directly or indirectly, the faintness of quiescent BH SXTs
to the lack of a hard surface in these systems.\footnote{As an
exception, if the quiescent X-ray emission of NS SXTs is powered by NS
cooling while coronal activity dominates in quiescent BH SXTs, the
lack of hard surface in BH SXTs is not crucial (but see objections to
the coronal emission scenario in \S4.2).} The observed luminosity
difference has therefore been interpreted as evidence for event
horizons in systems containing BH candidates.

This argument is constructed by elimination: to establish the presence
of event horizons in systems containing BH candidates, one should
eliminate the possibility of an exotic compact object with a hard
surface and a mass in excess of $2-3 M_\odot$ (the usual general
relativity limit for a NS, beyond which collapse to a BH is
traditionally unavoidable; Rhoades \& Ruffini 1974; Kalogera \& Baym
1996). In this section, a specific scenario which makes this exotic
assumption is considered in more detail (systems containing BH
candidates are still referred to as BH SXTs, for simplicity).

Assuming that the quiescent X-ray emission of SXTs is powered by
accretion,\footnote{The specific scenario considered here does not
exclude an additional contribution from NS cooling to the X-ray
emission of quiescent NS SXTs.}  the radiative efficiency should be
comparable in BH and NS SXTs (say $\sim 10 \%$), if hard surfaces are
present in both cases, because the compact objects should have
comparable compactness in first approximation (see below). Is it
possible for accretion to proceed at a rate $\sim 100$ times smaller in
BH SXTs during quiescence (only because of their $ \sim 5$ times more
massive compact objects) and thus explain the observed luminosity
difference?  As emphasized by Menou et al. (1999), the mass transfer
rates in BH and NS SXTs should be comparable, as long as one compares
systems with similar orbital periods. In addition, a close inspection
of the scaling with central mass of the accretion rate in quiescent
disks, in both the DIM and the layered accretion model of Menou
(2001), reveals only a weak dependence on the mass of the central
object, that is insufficient to explain the observed luminosity
difference. One is therefore tempted to focus on differences in the
properties of the flow in the vicinity of the compact object to
explain the observed luminosity difference.

\subsection{Accretion vs. Spindown Luminosity}

It is natural to expect BH candidates, even with a hard surface, to be
more compact than NSs, because of their larger masses. It is then
possible for NSs radii to exceed the radius of the marginally stable
orbit for an object of their mass, while BH candidates, being more
compact, could lie within their marginally stable orbit (currently
favored NS equations of state allow for NS radii both smaller and
larger than the corresponding marginally stable orbit radius;
e.g. Kalogera \& Baym 1996). Based on this difference between BH
candidates and NSs, an accretion scenario can be constructed in which
the quiescent X-ray luminosity difference is explained without
requiring BH candidates to have an event horizon.

For compact enough BH candidates, accretion will be transsonic and
transalfvenic because the flow radial speed approaches the speed of
light as the compact object radius approaches a Schwarzschild
radius. In this case, the gas reaching the BH candidate surface is
viscously disconnected from the rest of the flow and, independently of
the flow structure at large radii, emission from the gas shocking the
surface should have a radiative efficiency $\sim 10 \%$ or so. This
implies very low accretion rates, $\sim 10^{10-11}$~g~s$^{-1}$, in
quiescent BH SXTs.

It is reasonable to expect similar mass accretion rates in BH and NS
SXTs during quiescence (see point above). In the case of NS accretion,
however, if the NS radius exceeds the marginally stable orbit, the
entire flow is viscously connected. Medvedev \& Narayan (2001) have
shown that the flow can adopt a hot configuration in this case,
through which it effectively spins down the NS. Medvedev \& Narayan
(2001) note that the spindown luminosity dominates over the accretion
luminosity for low enough accretion rates. In particular, for
accretion rates $\sim 10^{10-11}$~g~s$^{-1}$, as discussed above, and
NS spin frequencies $\sim 300 $~Hz (a reasonable value for NSs in
SXTs; e.g., Wijnands \& van der Klis 1998; White \& Zhang 1997), a
spindown luminosity of $\sim 10^{33}$~erg~s$^{-1}$ is expected, much
in excess of the corresponding accretion luminosity. This spindown
luminosity is right at the level required to explain the quiescent
X-ray luminosity of NS SXTs (slight changes in model parameters, such
the NS spin rate, can easily account for the range of luminosities
observed in quiescent NS SXTs). Thus, in the accretion scenario
considered here, quiescent BH SXTs are less luminous than their NS
equivalents because, in their case, the flow is viscously disconnected
from the compact object and only the accretion luminosity, at a level
of $\sim 10^{30-31}$~erg~s$^{-1}$, is liberated (independently of the
BH candidate spin rate).  Note that the hot settling flow solution of
Medvedev \& Narayan (2001) is strictly valid only for an unmagnetized
NS. One can show that for surface field strengths roughly $\leq
10^8$~G (as is reasonable for NS in SXTs), an unmagnetized
approximation is relevant because the thermal pressure in the hot
settling flow dominates over the magnetic field pressure at the
stellar surface.

The accretion scenario outlined above is dynamically-consistent with
recent work on hot accretion flows and energetically-consistent with
the observed X-ray luminosity difference. It is arbitrary only in that
it postulates the existence of massive, very compact objects and it
assumes that NSs lie beyond their marginally stable orbits.  For
simplicity, the issue of thermal emission from the massive compact
object in BH SXTs and the role of magnetic fields associated with this
object were ignored (the fields could arguably be weak and the object
internal structure could be such that crustal heating like in the NS
case is not expected).

The value of this scenario may not be as a solid alternative to other
interpretations of the observed luminosity difference. Rather, it
shows that it could be dangerous to have evidence for BH event
horizons rely only on the luminosity difference between BH and NS
systems. By using additional observables, however, the evidence will
likely be made stronger. Detailed X-ray spectroscopic diagnostics may,
for instance, rule out the presence of a hot settling flow in
quiescent NS SXTs. Perhaps more importantly, accumulation of mass at
the hypothetical surface of the compact object in BH SXTs during
outburst may be expected to trigger type I X-ray bursts, by analogy
with the NS case (and independently of the detailed internal structure
of the hypothetical compact object). The non-detection of these bursts
may therefore constitute the strongest evidence for event horizons in
BH SXTs. Detailed burst calculations by Heyl \& Narayan (2002) for a
compact object of arbitrary size support the validity of this test.

\section*{Acknowledgments}

The author thanks Lars Bildsten, Jean-Pierre Lasota, Jeff McClintock,
Ramesh Narayan and Eliot Quataert for comments on the manuscript, and
the Center for Astrophysical Sciences at Johns Hopkins University for
hospitality.  Support for this work was provided by NASA through
Chandra Fellowship grant PF9-10006 awarded by the Smithsonian
Astrophysical Observatory for NASA under contract NAS8-39073.

\end{document}